\documentclass[aps,prl,groupedaddress,showpacs,nofootinbib]{revtex4}
\usepackage{graphicx}
\usepackage{times}
\usepackage{epsfig}
\include{epsf}
\usepackage{epstopdf}
\def\beq{\begin{equation}}
\def\eeq{\end{equation}}
\def\bey{\begin{eqnarray}}
\def\eey{\end{eqnarray}}

\def\lsim{\mathrel{\raise.3ex\hbox{$<$\kern-.75em\lower1ex\hbox{$\sim$}}}}
\def\gsim{\mathrel{\raise.3ex\hbox{$>$\kern-.75em\lower1ex\hbox{$\sim$}}}}

\begin{document}

\title{The Sensitivity of the IceCube Neutrino Detector to Dark Matter Annihilating in Dwarf Galaxies}
\author{Pearl Sandick$^{1}$, Douglas Spolyar$^{2,3}$, Matthew Buckley$^4$, Katherine Freese$^5$, and Dan Hooper$^{2,6}$
}
\address{
$^1$Theory Group and Texas Cosmology Center, The University of Texas at Austin, TX 78712 \\
$^2$Center for Particle Astrophysics, Fermi National Accelerator Laboratory, Batavia, IL  60510\\
$^3$Department of Physics, University of California,Santa Cruz, CA 95064   \\
$^4$Department of Physics, California Institute of Technology, Pasadena, CA 91125 \\
$^5$Michigan Center for Theoretical Physics, University of Michigan, Ann Arbor, MI 48109 \\
$^6$Department of Astronomy and Astrophysics, University of Chicago, Chicago, IL 60638 \\
}

\date{\today}

\begin{abstract}

In this paper, we compare the relative sensitivities of gamma-ray and neutrino observations
to the dark matter annihilation cross section in leptophilic models such as have been designed to 
explain PAMELA data.
We investigate whether the high energy neutrino telescope IceCube will be competitive with current and upcoming searches by gamma-ray telescopes, such as the Atmospheric \c{C}erenkov Telescopes (HESS, VERITAS and MAGIC), or the Fermi Gamma-Ray Space
Telescope, in detecting or constraining dark matter particles annihilating in dwarf spheroidal galaxies.  We find that after 10 years of observation of the most promising nearby dwarfs, 
IceCube will have sensitivity comparable to the current sensitivity of gamma-ray telescopes only for very heavy ($m_X \gtrsim 7$ TeV) or relatively light ($m_X \lesssim 200$ GeV) dark matter particles which annihilate primarily to $\mu^+\mu^-$.  If dark matter particles annihilate primarily to $\tau^+\tau^-$, IceCube will have superior sensitivity only for dark matter particle masses below the 200 GeV threshold of current Atmospheric \c{C}erenkov Telescopes.
If dark matter annihilations proceed directly to neutrino-antineutrino pairs a substantial fraction of the time, IceCube will be competitive with gamma-ray telescopes for a much wider range of dark matter masses.

\end{abstract}

\pacs{95.35.+d;95.30.Cq,98.52.Wz,95.55.Ka
\hspace{0.5cm} UTTG-13-09
\hspace{0.3cm} TCC-030-09
\hspace{0.3cm} FERMILAB-PUB-09-589-A}
\maketitle

\section{Introduction}

Of the many dark matter candidates to have been proposed, those within the classification of Weakly Interacting Massive Particles (WIMPs) are often considered to be the best motivated (for reviews, see Refs.~\cite{jungmanrev,bertonerev}). In particular, if a stable particle with an electroweak-scale mass and electroweak-scale interactions exists, such particles would have annihilated among themselves at a rate in the early universe that naturally led them to freeze-out with a relic density similar to the measured dark matter abundance. This process of dark matter annihilation is also predicted to be taking place in the present universe, providing the basis for dark matter indirect detection experiments. Such experiments search for the annihilation products of dark matter particles, including electrons and/or positrons, antiprotons, photons, and neutrinos. Promising sites for the observation of dark matter annihilation products include the core of the Sun~\cite{SOS}, the Earth~\cite{earth}, our Galactic halo~\cite{galaxy}, Galactic center~\cite{GC}, and dwarf satellite galaxies~\cite{dsph}. Here, we examine the sensitivity of the IceCube neutrino detector to dark matter annihilations taking place in dwarf satellite galaxies in the Milky Way.

In the case of dark matter candidates that annihilate primarily to gauge bosons or hadronic final states, gamma-ray telescopes provide a more sensitive test of dark matter annihilations taking place in dwarf spheroidals than can be accomplished with existing or planned neutrino telescopes.
 This is not necessarily the case, however, if the dark matter annihilates largely to leptons. Dark matter annihilating to leptonic final states has become increasingly well motivated by the anomalous observations recently reported by several cosmic ray experiments, including 
 PAMELA, which has observed an excess of cosmic ray positrons (relative to electrons) between 10 and 100 GeV~\cite{adriani} (along with previous experiments, including HEAT~\cite{heat} and AMS-01~\cite{ams01}, which also reported evidence for such an excess). A surplus of cosmic ray electrons and/or positrons has also been reported by the Fermi Gamma-Ray Space Telescope (FGST)~\cite{fgstelectrons}, however this excess is less than that previously reported by ATIC~\cite{chang}.

 Although an excess of cosmic ray positrons/electrons may come from more conventional sources such as pulsars~\cite{pulsars},
  a great deal of interest has been generated in the possibility that these signals might result from dark matter particles annihilating in the local halo of the Milky Way. Efforts to produce such signals with dark matter, however, have faced some model-building challenges. First, as there is no observed excess of cosmic ray antiprotons, the annihilation channels must be ``leptophilic," {\it i.e.}~the WIMPs must annihilate preferentially to leptons. Many models have been proposed with this property~\cite{leptophilic}. Second, the local halo density of dark matter is insufficient to explain the signals unless the annihilation rate is supplemented by a large factor $\sim 10 - 10^3$ relative to that predicted for a typical smoothly distributed thermal relic. Such a boost may arise due to the properties of the dark matter candidate itself; for example, via a nonperturbative Sommerfeld enhancement to the low-velocity cross section resulting from the exchange of a light state~\cite{sommerfeld}, or a Breit-Wigner enhancement~\cite{Feldman,Ibe:2008ye}.  Because of the smaller velocity dispersion of dark matter particles in smaller halos, the annihilation cross section in a dwarf galaxy may be even higher than that in the halo of the Milky Way~\cite{sommerfeld,kuhlen, strigari1,zentner}. Alternatively, an enhancement to the annihilation rate over the standard prediction can result if the dark matter is not distributed smoothly. In this paper, we examine bounds on the observed WIMP annihilation cross section, $\langle \sigma v \rangle$, including any relevant enhancements or boost factors, the origin of which we leave unspecified.

The prospects for IceCube observations of high energy neutrinos from dark matter annihilations or decays in the Galactic halo have been considered in~\cite{yuksel,ICann,ICdec,Mandal:2009yk}. It was found that IceCube will be sensitive to WIMP annihilation cross sections of order $10^{-24}$ to $10^{-23}$ cm$^{3}$s$^{-1}$, depending on the WIMP mass. As Super-Kamiokande is located in the Northern Hemisphere, and therefore is subject to significantly reduced atmospheric backgrounds from the direction of the Galactic center, its sensitivity to the neutrino flux from dark matter annihilations in the inner Milky Way has been studied in~\cite{meade,hisano} and found to be roughly an order of magnitude less sensitive than the IceCube projections.  Nonetheless, Ref.~\cite{meade} finds that Super-Kamiokande already constrains annihilations to $\tau^+ \tau^-$ as the explanation for the cosmic ray anomalies (though this channel is also disfavored by the large expected $\gamma$ flux), while Ref.~\cite{hisano} finds that Super-Kamiokande may also constrain the scenario in which neutrinos are directly produced in the annihilations as frequently as charged leptons.  While these studies are concerned with the signature of annihilations occurring in the Galactic center, here we focus instead on the annihilations taking place in satellite dwarf spheroidal galaxies of the Milky Way. In particular, we estimate the neutrino fluxes from WIMP annihilation in the dwarf galaxies Draco, Willman 1, and Segue 1, and examine the prospects for their observation with the IceCube neutrino telescope located at the South Pole.

\section{Dark Matter Annihilation In Dwarf Spheroidal Galaxies}

In general, the differential flux of particles of type $j$ from the annihilation of dark matter particles of mass $m_X$ is given by
 \begin{equation}
 \label{dN_dE}
 {d \Phi_j(\Delta \Omega, E_j) \over dE_j} = {\langle \sigma v \rangle \over 8 \pi m_X^2}
 \sum_F f_F  {dN_{j,F} \over dE_{j,F}} \times \bar J(\Delta \Omega) \Delta \Omega 
 \end{equation}
where $f_F$ is the fraction of annihilations which produce a final state $F$, and $dN_{j,F}/dE_{j,F}$ is the differential spectrum of particles $j$ from an annihilation to final state $F$.
$\bar J(\Delta \Omega)$ is the square of the dark matter density integrated along the line-of-sight, averaged over the solid angle $\Delta \Omega$:
 \begin{equation}
 \label{j}
 \bar{J}(\Delta \Omega) = {1 \over \Delta \Omega} \int_{\Delta \Omega} d\Omega \,
         \int_{l.o.s.} \rho_X^2(s) ds.
 \end{equation}

Dwarf spheroidal galaxies are promising sources for indirect dark matter searches for a number of reasons. First, they contain relatively large dark matter densities, and thus may produce sizable fluxes of annihilation products. Furthermore, dwarf spheroidals contain relatively little in the way of baryonic material (stars, gas, etc.) and are astrophysically simple.  The typical mass of known dwarf spheroidal galaxies is $\sim 10^7 \, {\rm M_\odot}$, which is distributed over a volume on the order of a cubic kiloparsec.  Dwarfs are largely devoid of astrophysical activity, and have very large mass to light ratios, Segue 1 being a particularly extreme example~\cite{geha}. The systems therefore lack astrophysical sources which could potentially mimic a signal of dark matter annihilation.  Although simulations suggest that many more are likely to exist, approximately 25 dwarf galaxies within the local group have been discovered thus far.

The dark matter distribution $\rho(r)$ within a dwarf spheroid can be fit with a five parameter density profile~\cite{strigari3} 
\begin{equation}
\label{genrho}
\rho(r) = \frac{\rho_0}{(\frac{r}{r_s})^a(1+(\frac{r}{r_s})^b)^\frac{c-a}{b}},
\end{equation}
where $r$ is the distance from the dynamical center of the dwarf galaxy, $r_s$ is the scale radius, and $\rho_0$ is the central core density. Typical ranges for the parameters $a$, $c$, and $b$, which determine the inner slope, outer slope, and transition between the two, respectively, are: $a=[0.0-1.5]$, $b=[0.5-1.5]$, and $c=[2-5]$.  $N$-body simulations find cusped inner profiles with $a=[1.0-1.5]$.  The profile of each individual halo can vary, however, depending upon its own merger history.  Given this uncertainty, we follow Ref.~\cite{strigari3} which marginalized over these five parameters, including $r_s$ and $\rho_0$, over the ranges specified above with flat priors. Although the parameter ranges are generous, Ref.~\cite{strigari3} finds that the velocity dispersion data fix the line-of-sight integral, $\bar{J}(\Delta \Omega)$, to lie in a relatively small range. For details, see Ref.~\cite{haloprifiles}.  

Presently, more than half a dozen dwarf galaxies are known which could potentially provide an observable flux of neutrinos.  These dwarfs are relatively near the Solar System (tens of kpc) and, importantly, are in the northern hemisphere. This enables IceCube's background of atmospheric muons to be largely avoided by looking through the Earth, towards the northern hemisphere. 
We have selected three nearby dwarfs as sources of interest for IceCube; Draco, Willman 1, and Segue 1.  In Table~\ref{tab:dwarfs}, we display the relevant properties of these satellite galaxies~\cite{ned}, with $\bar{J}(\Delta\Omega)\Delta\Omega$ and errors representing the $2\sigma$ range taken from Ref.~\cite{sehgalTeVPA}.
\begin{table}[h]
  \caption{Properties of selected Milky Way dwarf galaxies.
  \label{tab:dwarfs}}
  \begin{tabular}{l  c  c  c}
  Galaxy & Distance (kpc) & Log$_{10}\bar{J}(\Delta\Omega)\Delta\Omega$ & Declination \\
  \hline
  Draco     & $80$  & $18.63 \pm 0.73$ & $+57^\circ 54' 55"$    \\
  Willman 1 & $38$  & $19.55 \pm 1.19$ & $+51^\circ 03' 00"$  \\
  Segue 1   & $23$  & $19.88 \pm 0.82$ & $+16^\circ 04' 55"$
  \end{tabular}
\end{table}
We note that there is some uncertainty in the observational status of Willman 1; it is currently unclear whether it is a dwarf galaxy or a globular cluster.  It follows that there is considerably more uncertainty in the value of $\bar{J}(\Delta\Omega)\Delta\Omega$.

IceCube is designed to have an angular resolution of approximately $1^\circ$ for muon tracks~\cite{resconi}.  Given this resolution, the dwarf galaxies in Table~\ref{tab:dwarfs} are sufficiently small and distant that we can treat them as point sources.  Atmospheric \c{C}erenkov Telescopes and the Fermi Gamma Ray Space Telescope have angular resolutions much smaller than IceCube. For example, VERITAS has observed Draco, Ursa Minor, and Willman 1 over regions within $0.15^\circ$ of each galaxy's center~\cite{hui}.


Several dwarf galaxies have been observed by ACTs as
potential sources of gamma-rays:
Draco by MAGIC~\cite{albert} and STACEE~\cite{stacee}, Sagittarius (which is visible from
the Southern Hemisphere) by H.E.S.S.~\cite{aharonian}, and  Draco, Willman 1, and Ursa Minor by VERITAS~\cite{hui}.   None of these telescopes have observed any significant signal and, therefore, have placed upper limits on the flux of gamma-rays coming from these sources. The strongest constraints on the annihilation rate of WIMPs in dwarf galaxies come from H.E.S.S. and VERITAS~\cite{strigari3}.   H.E.S.S. observations of the center of the Sagittarius dwarf yield an upper limit on the gamma-ray flux above 250 GeV of  
$\Phi(E_{\gamma}>{\rm 250 \,GeV}) \lsim 3.6 \times 10^{-12} \,{\rm cm^{-2} s^{-1}}$
at the 95\% confidence level~\cite{aharonian}. 
VERITAS observations of Draco, Ursa Minor, and Willman 1 bound the gamma-ray flux from each of these objects above 200~GeV to be $\Phi(E_{\gamma}>{\rm 200 \,GeV}) \lsim 2.4 \times 10^{-12} \,{\rm cm^{-2} s^{-1}}$~\cite{hui}.


\section{Muon Neutrinos From Dwarf Spheroidal Galaxies}

In this study, we consider three leptophilic scenarios: WIMPs which annihilate solely to $\tau^+ \tau^-$, solely $\mu^+ \mu^-$, or to both $\mu^+ \mu^-$ and $\nu_\mu \bar{\nu}_\mu$, each with branching fractions of 50\%. 
The differential flux of neutrinos from annihilations directly to neutrino-antineutrino pairs in a dwarf galaxy is given by Eq.~\ref{dN_dE}, and the $\nu_\mu$  differential spectrum at production takes the very simple form
\beq
\label{nuspectrum}
\frac{dN_{\nu_\mu}}{dE_{\nu_\mu}}=\frac{dN_{\bar{\nu}_\mu}}{dE_{\bar{\nu}_\mu}}=\delta(E_{\nu_\mu}-m_X).
\eeq
In the cases in which the WIMPs annihilate to $\tau^+\tau^-$ or $\mu^+\mu^-$,  we use PYTHIA~\cite{pythia} to calculate the resulting neutrino spectra.  In all cases, we include the effects of three-flavor vacuum oscillations~\cite{pdg}.


Atmospheric muon neutrinos constitute the most serious background for detection of $\nu_\mu$ from dark matter annihilation products from northern hemisphere dwarf spheroidal galaxies. This background is a function of the zenith angle observed~\cite{gaisser}. We use Ref.~\cite{Honda:2006qj} to calculate the atmospheric neutrino background specific to the direction of each of the three source dwarfs.  We note that the predicted background spectra are in agreement with the observations of Amanda-II~\cite{Collaboration:2009nf}.


The rate of muon tracks from charged current neutrino interactions observed at IceCube is calculated by combining the incoming spectrum of muon neutrinos with the probability of those neutrinos being converted to muons above the energy threshold of the telescope~\cite{halzen}, which is given by
\begin{equation}
\label{nuprob}
P(E_\nu, E^{{\rm thr}}_\mu)_{\mu}=\rho \,\, {\rm N_A} \int^1_0 dy \frac{d\sigma_{\nu N}}{dy}(E_\nu,y) \, [R_{\mu}(E_\nu (1-y),E^{{\rm thr}}_\mu)+L].
\end{equation}
Here, $\rho=\rho_{\rm ice}\approx0.9$  g cm$^{-3}$ is the density of the target medium (ice)
and $\rm N_A=6.022\times10^{23}$ is Avogadro's number.  The total charged current
cross section for neutrino nucleon scattering, $\sigma_{\nu N}$, can be taken as
\beq
\frac{\sigma_{\nu N}(E_\nu)+\sigma_{\bar\nu N}(E_\nu)}{2}=3.06\times10^{-36} \bigg(\frac{E_\nu}{600 \,{\rm GeV}}\bigg)^{0.96} {\rm cm^{2}}
\eeq
for 100 GeV $< E_\nu < 1$ TeV, and is approximately flatly distributed in $y$~\cite{gandhi}. Finally, $R_{\mu}+L$ is the effective detector size, the sum of the physical length of the detector, $L$, and the distance a muon 
travels before its energy falls below the threshold of the experiment, $R_\mu$.  The muon range in ice is 
\begin{equation}
R_{\mu} \approx 2.4\,{\rm km} \times \ln\bigg[\frac{2.0+4.2 (E_{\mu}/{\rm TeV})}{2.0+4.2 (E_\mu^{{\rm thr}}/{\rm TeV})}\bigg].
\end{equation}
The event rate for both the background and the signal is given by the integral
\begin{equation}
\label{events}
{N}=\int A_{\rm eff} \, P(E_\nu,E_\mu^{{\rm thr}})\frac{d\Phi}{dE_\nu} dE_\nu.
\end{equation}
The effective area, $A_{\rm eff}$, in IceCube is approximately one square kilometer. The DeepCore supplement to the 
IceCube detector is designed to increase the sensitivity to low energy neutrino-induced muons with 10 GeV $\lesssim 
E_\mu \lesssim$ 
100 GeV, with the improvement most substantial at the lowest energies~\cite{nuArea}.  However, the dark matter 
annihilation signal in IceCube and DeepCore is dominated by muons with energies close to the dark matter mass: for the lowest WIMP masses considered here, $m_X 
=100$ GeV, the signal comes primarily from neutrinos with 50 GeV $\lesssim E_\nu \lesssim$ 100 GeV.  In this 
case, considering only the IceCube and DeepCore sensitivities reported in Fig. 4 of 
Ref.~\cite{nuArea}, we expect an improvement of roughly 20\% over IceCube alone.  

WIMP masses below $\sim 100$ GeV are not compatible with the anomalous cosmic ray excesses.  However, the sensitivity 
to neutrinos with $E_\nu \sim 10$ GeV is improved by including the DeepCore addition by more than an order of 
magnitude over that of IceCube alone~\cite{nuArea}.  At the same time, since the maximal angle between a 
muon track and the primary neutrino increases at low neutrino energies as $1/\sqrt{E_\nu}$, the angular resolution is 
significantly degraded at low energies, resulting in a decreased signal to background ratio for localized objects such 
as the dwarf satellite galaxies in Table~\ref{tab:dwarfs}. We therefore estimate that the DeepCore addition would 
improve the sensitivity to low-mass dark matter annihilating in dwarf satellite galaxies by, at most, a factor of a few.  
However, as the cosmic ray anomalies do not favor light leptophilic dark matter, we report only on the 
sensitivity of the IceCube neutrino detector to dark matter annihilations in dwarf galaxies for WIMPs with $m_X > 100$ 
GeV.

In comparing the event rate to that predicted from the background of atmospheric neutrinos, we include events with energy up to 25\% greater than the WIMP mass, which conservatively accounts for the finite energy resolution of IceCube~\cite{resconi}.
We can now calculate the exposure time, T, necessary to achieve a detection with a given level of statistical significance over the background of atmospheric neutrinos.
Since the background is essentially isotropic over the angular window of interest we follow the analysis of Ref.~\cite{bergstrom} 
to find the  minimum exposure time,
\begin{equation}
{\rm T}= \sigma_{stat}^2 \frac{N_{\rm atm}+N_{\rm DMA}}{N_{\rm DMA}^2},
\end{equation}
where $N_{\rm DMA}$ is the event rate from dark matter annihilations in 
the dwarf and $N_{\rm atm}$ is the event rate from the atmospheric background. 
In the following section, we use this approach to determine the sensitivity of IceCube to dark matter annihilations taking place in the dwarf galaxies Draco, Willman 1, and Segue 1.

\section{Prospects For IceCube And Comparison To Gamma-Ray Searches}

\begin{figure}[t]
\begin{center}
\mbox{\includegraphics[width=.47\textwidth]{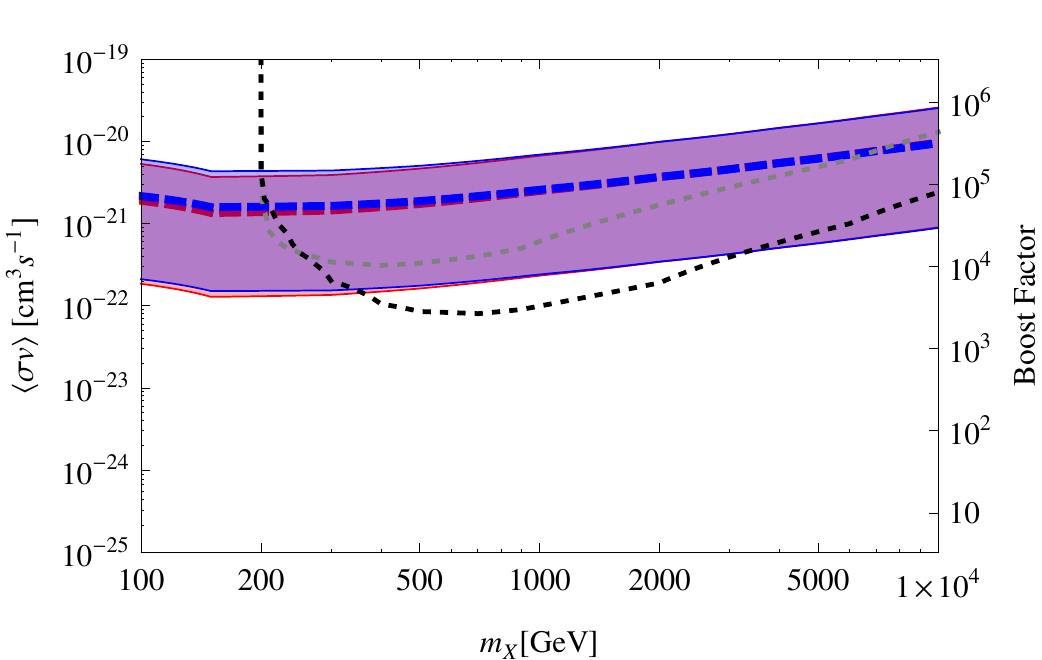}}
\hspace{0.03\textwidth}
\mbox{\includegraphics[width=.47\textwidth]{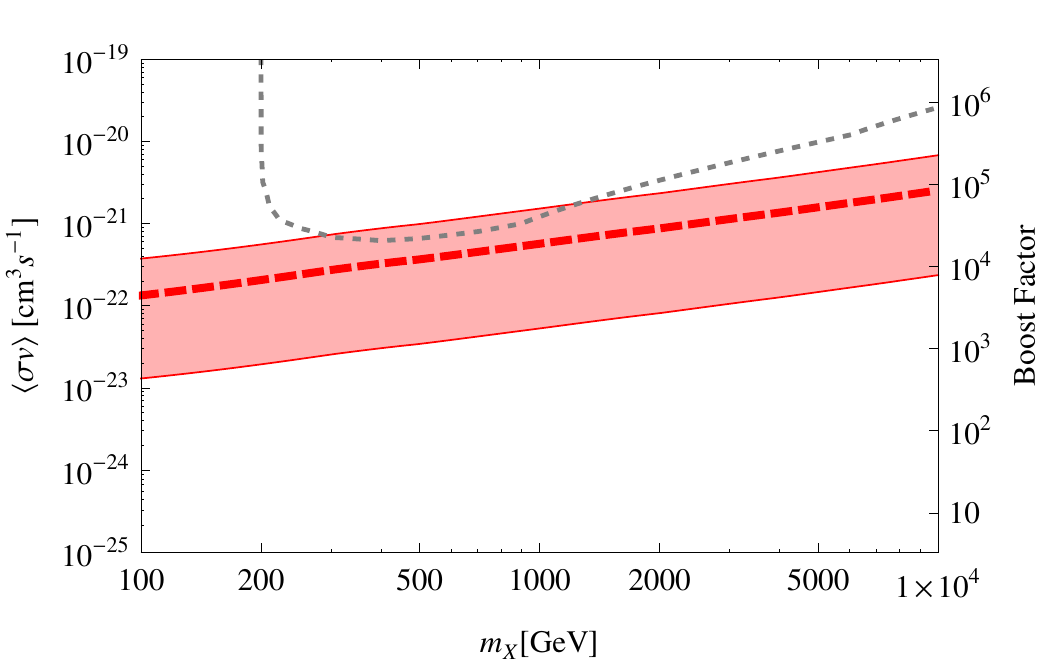}}
\end{center}
\vspace{-0.5cm}
\caption{\it Projected upper limits on the dark matter annihilation cross section, $\langle \sigma v \rangle$ (alternatively, boost factor), at 95\% confidence level as a function of mass (thick dashed) from neutrino observations of the dwarf spheroidal galaxy Draco. Also shown are the 5$\sigma$ discovery regions (shaded). In the left frame, the red (lower) region and lines denote the case in which dark matter annihilates to $\mu^+\mu^-$, whereas  the blue (upper) regions and lines denote annihilations to $\tau^+\tau^-$. The right frame is for the case in which dark matter annihilates to both $\mu^+\mu^-$ and  $\nu_{\mu}\bar{\nu}_{\mu}$ with equal probability. In each frame, the dotted lines denote the current limits from gamma-ray observations by VERITAS. See text for more details.
\label{fig:draco}}
\end{figure}

We now compare the relative merits of gamma-ray and neutrino observations
in their ability to place limits on the dark matter annihilation cross section in leptophilic models. 
In Fig.~\ref{fig:draco}, we present IceCube's sensitivity to neutrinos from dark matter annihilations in Draco, after 10 years of observation. In the left panel, we show results for the annihilation channels $XX \rightarrow \mu^+ \mu^-$ in red (lower contours) and $XX \rightarrow \tau^+ \tau^-$ in blue (upper contours).  As a consequence of neutrino oscillations, red and blue contours are distinct only at low $m_X$. The thick dashed lines denote the projected 2$\sigma$ upper limit on the annihilation cross section as a function of mass (using the $2\sigma$ low value for $\bar{J}(\Delta\Omega)\Delta\Omega$), whereas the shaded regions bounded by thin solid lines denote 
the minimal cross sections necessary for a 5$\sigma$ discovery, given the $\pm 2\sigma$ range of values for $\bar{J}(\Delta\Omega)\Delta\Omega$. 
In the right panel, results are shown for the case in which $XX \rightarrow \mu^+ \mu^-$ and $XX \rightarrow \nu_\mu \bar{\nu}_\mu$ each occur with branching fractions of 50\%. One can see that, due to the uncertainty in the dark matter halo properties, the potential discovery regions extend well below the exclusion contours. 

For comparison, in each frame we also show the corresponding upper limits on the dark matter annihilation cross section from VERITAS observations of Draco, as found in Ref.~\cite{strigari3}.  In the left panel, the grey and black dotted curves represent the upper limits on $\langle \sigma v \rangle$ assuming annihilations to $\mu^+ \mu^-$ and $\tau^+ \tau^-$, respectively, while in the right panel, the grey dotted curve is the VERITAS upper limit assuming $X X \rightarrow \mu^+ \mu^-$ and $XX \rightarrow \nu_\mu \bar{\nu}_\mu$ with equal branching fractions. We note that Ref.~\cite{strigari3} uses the 90\% confidence level lower limit on $\bar{J}(\Delta\Omega)\Delta\Omega$, while we consider the slightly more generous $2\sigma$ lower limit.  Our projected sensitivities are therefore slightly more conservative. As Draco is the least bright of the dwarf galaxies considered here, we find that for WIMP masses $\gtrsim 200$ GeV, VERITAS has already constrained the annihilation cross section roughly as tightly as IceCube will with 10 years of data for annihilation purely to $\mu^+ \mu^-$ or $\tau^+ \tau^-$.  In the case of annihilations to $\mu^+ \mu^-$, IceCube may place competitive limits on the WIMP annihilation cross section only for $m_X  \gsim 5$ TeV or $m_X \lesssim 200$ GeV.  For annihilation to $\tau^+ \tau^-$ the current constraints on $\langle \sigma v \rangle$ from VERITAS observations of Draco are stronger than any IceCube will be able to achieve in the next 10 years for $m_X > 200$ GeV.  This stems from the fact that taus often decay hadronically, resulting in many more high energy photons than in the purely leptonic decays of muons. For annihilation to taus, a substantial portion of IceCube's discovery reach has already been excluded by VERITAS.

Instead of presenting our results as limits (or discovery reach) on the dark matter annihilation cross section, we can alternatively describe them in terms of the boost factor to the annihilation rate in a given dwarf spheroidal galaxy. Along the right side of each frame, the constraints and discovery prospects are given in terms of this quantity, which we define as the annihilation rate divided by that predicted for dark matter with $\langle \sigma v \rangle = 3 \times 10^{-26}$ cm$^3$/s and distributed smoothly over the halo. This definition allows us to leave the origin of the boost factor unspecified.

Turning to the right panel, we find that if neutrinos and muons are each are directly produced in 50\% of annihilations, IceCube will have comparable or better sensitivity than VERITAS for all WIMP masses.
In fact, in this case IceCube's discovery reach is largely untested by ACTs. 
We note that if the branching fraction to neutrinos is larger than 50\%, IceCube's sensitivity will be further increased, while that of ACTs and other gamma-ray telescopes will be reduced.

\begin{figure}[t]
\begin{center}
\mbox{\includegraphics[width=.47\textwidth]{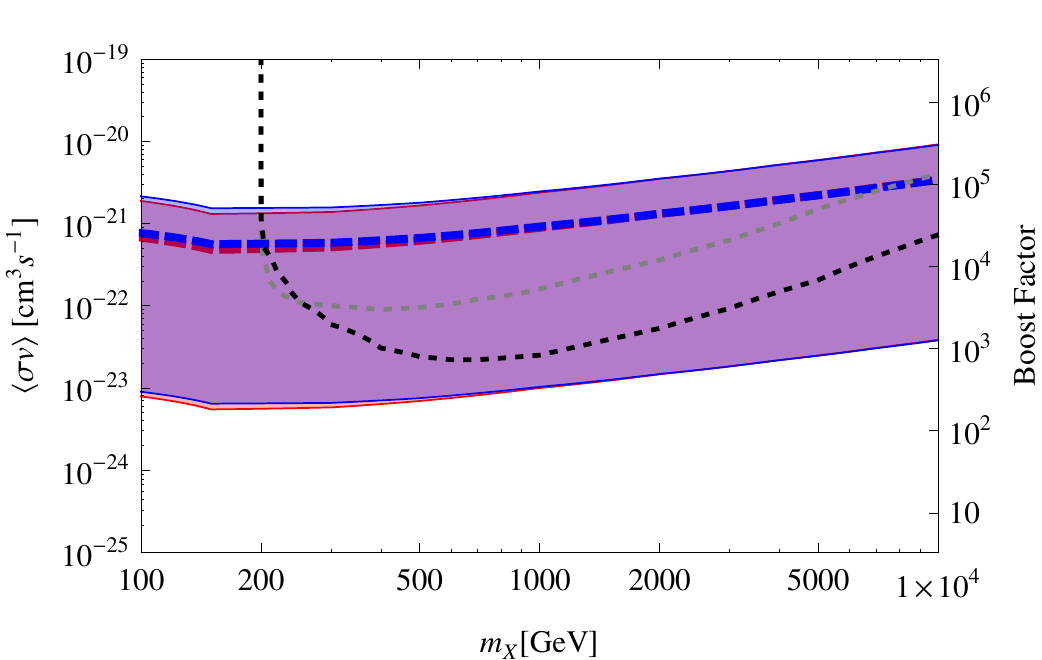}}
\hspace{0.03\textwidth}
\mbox{\includegraphics[width=.47\textwidth]{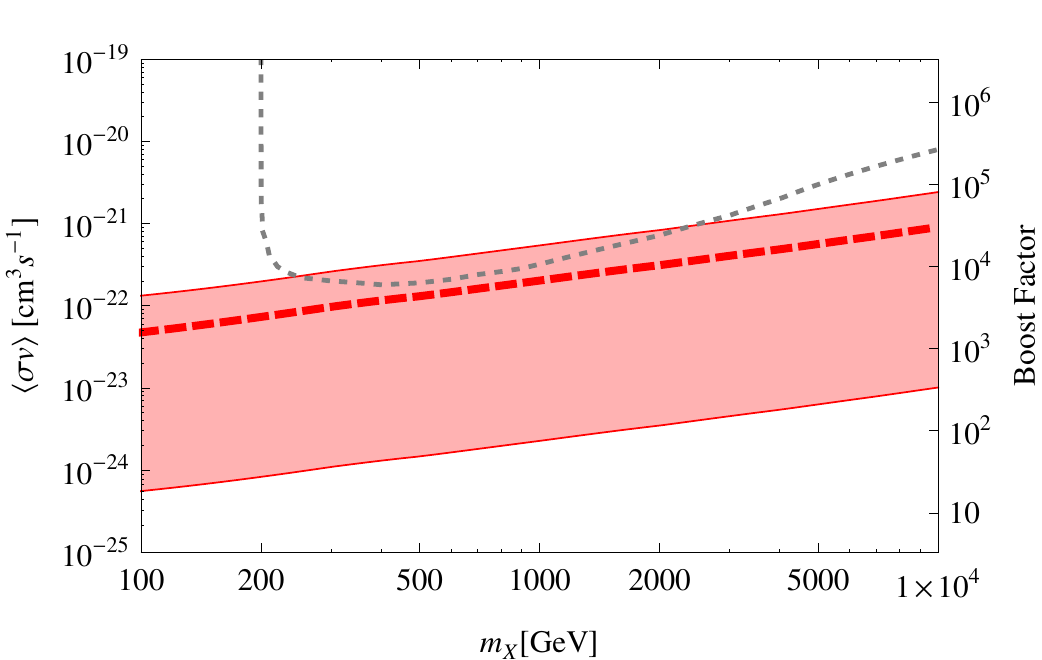}}
\end{center}
\vspace{-0.5cm}
\caption{\it The same as Fig.~1, but for observations of the dwarf spheroidal galaxy Willman 1.  
\label{fig:willman}}
\end{figure}

In Fig.~\ref{fig:willman} we present the same information as Fig.~\ref{fig:draco}, but for the dwarf galaxy Willman 1.  Willman 1 is somewhat brighter than Draco, and therefore we expect all sensitivities to improve.  However, there is also considerably more uncertainty in the halo profile, leading to large discovery regions and inflated prospective limits.  For annihilations to charged leptons only, boost factors of ${\cal O}(10^2)$ may be accessible at IceCube if $\bar{J}(\Delta\Omega)\Delta\Omega$ is near its $2\sigma$ upper limit. For low $m_X$, these annihilation cross sections have not yet been probed by ACTs.  For annihilation to $\mu^+\mu^-$ and $\nu_\mu \bar{\nu}_\mu$, shown in the right panel, boost factors as low as ${\cal O}(10)$ may be accessible in the most optimistic case.

\begin{figure}[t]
\begin{center}
\mbox{\includegraphics[width=.47\textwidth]{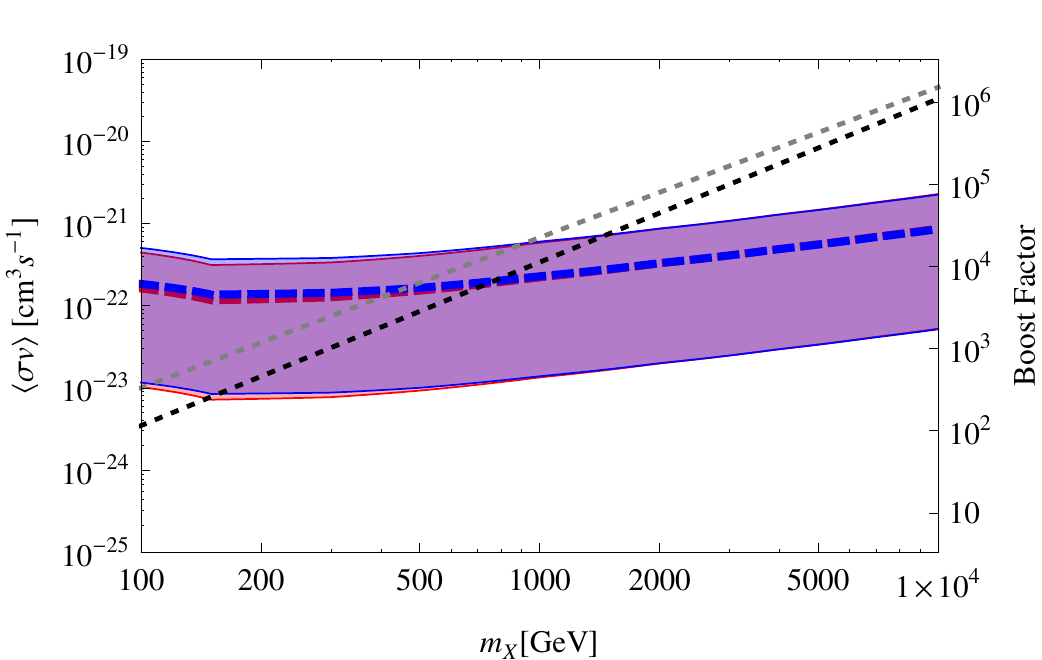}}
\hspace{0.03\textwidth}
\mbox{\includegraphics[width=.47\textwidth]{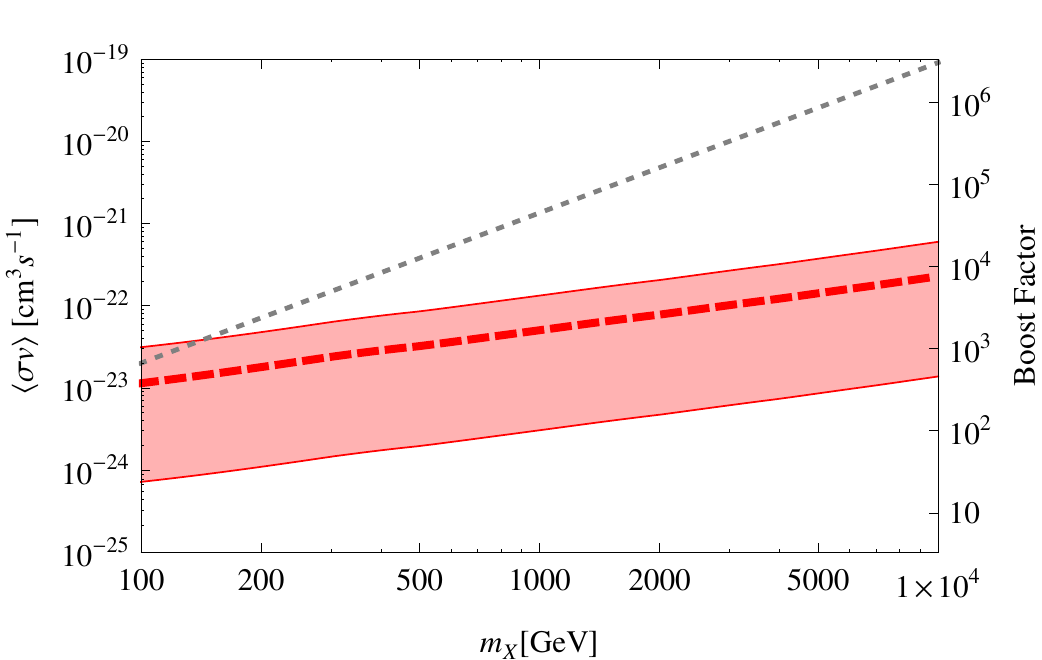}}
\end{center}
\vspace{-0.5cm}
\caption{\it The same as Fig.~1, but for observations of the dwarf spheroidal galaxy Segue 1. In this figure, the dotted lines denote the preliminary limits from the Fermi Gamma-Ray Space Telescope~\cite{fgst}.
\label{fig:segue}}
\end{figure}

Finally, we examine the closest of the three dwarfs we have considered, Segue 1.  In Fig.~\ref{fig:segue}, we present IceCube's projected sensitivity to dark matter annihilations in Segue 1, along with the current upper limits on the annihilation cross section from FGST measurements~\cite{fgst}.  For annihilation to muons (taus), IceCube may eventually set stronger exclusion limits for $m_X > 450 (750)$ GeV. If there is a 50\% branching fraction to neutrinos, IceCube will do better for all WIMP masses shown. As Segue 1 is newly discovered, we expect that ACTs will observe this galaxy in the near future.


\section{Discussion and Conclusions}

In this article, we have considered the prospects for the kilometer-scale neutrino telescope IceCube to detect neutrinos from dark matter annihilating in dwarf spheroidal galaxies, and have compared these to the constraints placed by gamma-ray telescopes. We find that if dark matter annihilates primarily to muons, taus, and/or neutrinos, IceCube can potentially provide constraints comparable to or even stronger than those obtained by current gamma-ray telescopes.

Other dwarf spheroidal galaxies, such as Ursa Minor, which yield weaker gamma-ray limits on the dark matter annihilation cross section will also be less promising to be observed by IceCube.  An analysis of the neutrino flux from all dwarf galaxies accessible to IceCube could potentially improve the sensitivity to the annihilation cross section, however, the maximal improvement assuming all dwarfs are equal is proportional to $1/\sqrt{N}$, where $N$ is the number of dwarfs in the sample. All dwarfs, of course, are not equal.  In order to carry out such an analysis, one must assume that the mechanism responsible for the boosted (relative to thermal) annihilation cross section results in the same observed cross section in each dwarf galaxy.  As the velocity dispersion in each dwarf is independent and the clumpiness of each dwarf is unknown, there is no reason to expect that the observed annihilation cross section in any two dwarfs should be the same, thus, at this point, the only appropriately model-independent interpretation is to view the annihilation signal from each dwarf independently. Given the velocity dispersion in each dwarf, one could derive limits on particular dark matter models assuming a specific velocity-dependent enhancement to the annihilation cross section. Such a study may be useful in the near future.

\bigskip 

KF is supported by the US Department of Energy and MCTP via the Univ.~of Michigan and the National Science Foundation under Grant No.~PHY-0455649; DS and DH are supported by the US Department of Energy, including grant DE-FG02-95ER40896; DH is also supported by NASA grant NAG5-10842; PS is supported by the National Science Foundation under Grant No.~PHY-0455649. MB is supported by the US Department of Energy, under grant DE-FG03-92-ER40701. K. F. would like to thank the Aspen
Center for Physics and the Texas Cosmology Center, and P. S. would like to thank MCTP.


\begin{thebibliography}{}

\bibitem{jungmanrev}
  G.~Jungman, M.~Kamionkowski and K.~Griest,
  Phys.\ Rept.\  {\bf 267}, 195 (1996)
  [arXiv:hep-ph/9506380].

\bibitem{bertonerev}
  G.~Bertone, D.~Hooper and J.~Silk,
  Phys.\ Rept.\  {\bf 405}, 279 (2005)
  [arXiv:hep-ph/0404175].

\bibitem{SOS}
  M.~Srednicki, K.~A.~Olive and J.~Silk,
  Nucl.\ Phys.\  B {\bf 279}, 804 (1987).

\bibitem{earth}
  K.~Freese,
  Phys.\ Lett.\  B {\bf 167}, 295 (1986);
%
  L.~M.~Krauss, M.~Srednicki and F.~Wilczek,
  Phys.\ Rev.\  D {\bf 33}, 2079 (1986).

\bibitem{galaxy}
  J.~R.~Ellis, R.~A.~Flores, K.~Freese, S.~Ritz, D.~Seckel and J.~Silk,
  Phys.\ Lett.\  B {\bf 214}, 403 (1988);
%
  M.~S.~Turner and F.~Wilczek,
  Phys.\ Rev.\  D {\bf 42}, 1001 (1990);
%
  M.~Kamionkowski and M.~S.~Turner,
  Phys.\ Rev.\  D {\bf 43}, 1774 (1991);
%
  J.~Silk and M.~Srednicki,
  Phys.\ Rev.\ Lett.\  {\bf 53}, 624 (1984).

\bibitem{GC}
  L.~Bergstrom, P.~Ullio and J.~H.~Buckley,
  Astropart.\ Phys.\  {\bf 9}, 137 (1998)
  [arXiv:astro-ph/9712318].
%
 

\bibitem{dsph}
  N.~W.~Evans, F.~Ferrer and S.~Sarkar,
  Phys.\ Rev.\  D {\bf 69}, 123501 (2004)
  [arXiv:astro-ph/0311145];
  L.~Bergstrom and D.~Hooper,
  Phys.\ Rev.\  D {\bf 73}, 063510 (2006)
  [arXiv:hep-ph/0512317].



\bibitem{yuksel}
  H.~Yuksel, S.~Horiuchi, J.~F.~Beacom and S.~Ando,
  Phys.\ Rev.\  D {\bf 76}, 123506 (2007)
  [arXiv:0707.0196 [astro-ph]].

\bibitem{ICann}
  D.~Spolyar, M.~Buckley, K.~Freese, D.~Hooper and H.~Murayama,
  arXiv:0905.4764 [astro-ph.CO].

\bibitem{ICdec}
  M.~R.~Buckley, K.~Freese, D.~Hooper, D.~Spolyar and H.~Murayama,
  arXiv:0907.2385 [astro-ph.HE].

\bibitem{Mandal:2009yk}
  S.~K.~Mandal, M.~R.~Buckley, K.~Freese, D.~Spolyar and H.~Murayama,
  arXiv:0911.5188 [hep-ph].

\bibitem{meade}
  P.~Meade, M.~Papucci, A.~Strumia and T.~Volansky,
  arXiv:0905.0480 [hep-ph].

\bibitem{hisano}
  J.~Hisano, M.~Kawasaki, K.~Kohri and K.~Nakayama,
  Phys.\ Rev.\  D {\bf 79}, 043516 (2009)
  [arXiv:0812.0219 [hep-ph]].

\bibitem{adriani}
  O.~Adriani {\it et al.}  [PAMELA Collaboration],
  Nature {\bf 458}, 607 (2009)
  [arXiv:0810.4995 [astro-ph]].

\bibitem{heat}
  S.~W.~Barwick {\it et al.}  [HEAT Collaboration],
  Astrophys.\ J.\  {\bf 482}, L191 (1997)
  [arXiv:astro-ph/9703192];
%
  S.~Coutu {\it et al.},
  {\it Prepared for 27th International Cosmic Ray Conference (ICRC 2001), Hamburg, 
  Germany, 2001}, IUPAP, London, p. 1687.

\bibitem{ams01}
  M.~Aguilar {\it et al.}  [AMS-01 Collaboration],
  Phys.\ Lett.\  B {\bf 646}, 145 (2007)
  [arXiv:astro-ph/0703154].


\bibitem{fgstelectrons}
  A.~A.~Abdo {\it et al.}  [The Fermi LAT Collaboration],
  Phys.\ Rev.\ Lett.\  {\bf 102}, 181101 (2009)
  [arXiv:0905.0025 [astro-ph.HE]];
  D.~Grasso {\it et al.}  [FERMI-LAT Collaboration],
  Astropart.\ Phys.\  {\bf 32}, 140 (2009)
  [arXiv:0905.0636 [astro-ph.HE]].

\bibitem{chang}
  J.~Chang {\it et al.},
  Nature {\bf 456}, 362 (2008).

\bibitem{pulsars}
  D.~Hooper, P.~Blasi and P.~D.~Serpico,
  JCAP {\bf 0901}, 025 (2009)
  [arXiv:0810.1527 [astro-ph]];
  H.~Yuksel, M.~D.~Kistler and T.~Stanev,
  Phys.\ Rev.\ Lett.\  {\bf 103}, 051101 (2009)
  [arXiv:0810.2784 [astro-ph]];
  S.~Profumo,
  arXiv:0812.4457 [astro-ph].


\bibitem{leptophilic}
  R.~Harnik and G.~D.~Kribs,
  Phys.\ Rev.\  D {\bf 79}, 095007 (2009)
  [arXiv:0810.5557 [hep-ph]];
%
  A.~E.~Nelson and C.~Spitzer,
  arXiv:0810.5167 [hep-ph];
%
  I.~Cholis, D.~P.~Finkbeiner, L.~Goodenough and N.~Weiner,
  arXiv:0810.5344 [astro-ph];
%
  K.~M.~Zurek,
  Phys.\ Rev.\  D {\bf 79}, 115002 (2009)
  [arXiv:0811.4429 [hep-ph]];
%
  P.~J.~Fox and E.~Poppitz,
  Phys.\ Rev.\  D {\bf 79}, 083528 (2009)
  [arXiv:0811.0399 [hep-ph]];
%
  C.~R.~Chen and F.~Takahashi,
  JCAP {\bf 0902}, 004 (2009)
  [arXiv:0810.4110 [hep-ph]];
%
  I.~Cholis, G.~Dobler, D.~P.~Finkbeiner, L.~Goodenough and N.~Weiner,
  arXiv:0811.3641 [astro-ph];
%
  D.~Hooper and K.~M.~Zurek,
  Phys.\ Rev.\  D {\bf 79}, 103529 (2009)
  [arXiv:0902.0593 [hep-ph]];
%
  D.~Hooper and T.~M.~P.~Tait,
Phys.\ Rev.\ D, in press, arXiv:0906.0362 [hep-ph];
%
  N.~Arkani-Hamed, D.~P.~Finkbeiner, T.~R.~Slatyer and N.~Weiner,
  Phys.\ Rev.\  D {\bf 79}, 015014 (2009)
  [arXiv:0810.0713 [hep-ph]].


\bibitem{sommerfeld}
  J.~March-Russell, S.~M.~West, D.~Cumberbatch and D.~Hooper,
  JHEP {\bf 0807}, 058 (2008)
  [arXiv:0801.3440 [hep-ph]];
  M.~Cirelli and A.~Strumia,
  arXiv:0808.3867 [astro-ph].
  
\bibitem{Feldman}
  D.~Feldman, Z.~Liu and P.~Nath,
  Phys.\ Rev.\  D {\bf 79}, 063509 (2009)
  [arXiv:0810.5762 [hep-ph]].

\bibitem{Ibe:2008ye}
  M.~Ibe, H.~Murayama and T.~T.~Yanagida,
  Phys.\ Rev.\  D {\bf 79}, 095009 (2009)
  [arXiv:0812.0072 [hep-ph]].

\bibitem{kuhlen}
  M.~Kuhlen, J.~Diemand and P.~Madau,
  Astrophys.\ J.\ {\bf 686}, 262 (2008)
  [arXiv:0805.4416 [astro-ph]].

\bibitem{strigari1}
  L.~E.~Strigari, S.~M.~Koushiappas, J.~S.~Bullock and M.~Kaplinghat,
  Phys.\ Rev.\  D {\bf 75}, 083526 (2007)
  [arXiv:astro-ph/0611925].

\bibitem{zentner}
  B.~Robertson and A.~Zentner,
  Phys.\ Rev.\  D {\bf 79}, 083525 (2009)
  [arXiv:0902.0362 [astro-ph.CO]].


\bibitem{geha}
  M.~Geha, B.~Willman, J.~D.~Simon, L.~E.~Strigari, E.~N.~Kirby, D.~R.~Law and J.~Strader,
  Astrophys.\ J.\  {\bf 692}, 1464 (2009)
  [arXiv:0809.2781 [astro-ph]].

\bibitem{strigari3}
  R.~Essig, N.~Sehgal and L.~E.~Strigari,
  Phys.\ Rev.\  D {\bf 80}, 023506 (2009)
  [arXiv:0902.4750 [hep-ph]].

\bibitem{haloprifiles}
  L.~E.~Strigari, J.~S.~Bullock, M.~Kaplinghat, J.~D.~Simon, M.~Geha, B.~Willman and M.~G.  ~Walker,
  Nature {\bf 454}, 1096 (2008)
  [arXiv:0808.3772 [astro-ph]].

\bibitem{ned}
  NASA/IPAC Extragalactic Database, 
  November 24, 2009, 
  http://nedwww.ipac.caltech.edu/

\bibitem{sehgalTeVPA}
  Neelima Sehgal, talk presented at TeV Particle Astrophysics, 
  July 14, 2009,  
  http://www-conf.slac.stanford.edu/tevpa09/Sehgal090714.pdf 

\bibitem{resconi}
  E.~Resconi for the Icecube Collaboration,
  Nucl.\ Instrum.\ Meth.\  A {\bf 602}, 7 (2009)
  [arXiv:0807.3891 [astro-ph]].

\bibitem{hui}
  C.~M.~Hui (for the VERITAS Collaboration),
  AIP Conf.\ Proc.\  {\bf 1085}, 407 (2009)
  [arXiv:0810.1913 [astro-ph]].

\bibitem{albert}
  J.~Albert {\it et al.}  [MAGIC Collaboration],
  Astrophys.\ J.\  {\bf 679}, 428 (2008)
  [arXiv:0711.2574 [astro-ph]].

\bibitem{stacee}
  D.~D.~Driscoll {\it et al.}  [STACEE Collaboration],
  arXiv:0710.3545 [astro-ph].

\bibitem{aharonian}
  :.~F.~Aharonian  [HESS Collaboration],
  Astropart.\ Phys.\  {\bf 29}, 55 (2008)
  [arXiv:0711.2369 [astro-ph]].

\bibitem{pythia}
T.~Sjostrand, {\it et al.},
Comput. Phys. Commun., 135, 238 (2001).

\bibitem{pdg}
  C.~Amsler {\it et al.}  [Particle Data Group],
  Phys.\ Lett.\  B {\bf 667}, 1 (2008).

  D.~Grellscheid and P.~Richardson,
  arXiv:0710.1951 [hep-ph].

\bibitem{gaisser}
  T.~K.~Gaisser and M.~Honda,
  Ann.\ Rev.\ Nucl.\ Part.\ Sci.\  {\bf 52}, 153 (2002)
  [arXiv:hep-ph/0203272].

\bibitem{Honda:2006qj}
  M.~Honda, T.~Kajita, K.~Kasahara, S.~Midorikawa and T.~Sanuki,
  Phys.\ Rev.\  D {\bf 75}, 043006 (2007)
  [arXiv:astro-ph/0611418];
  see also, G. Barr {\it et al}, Phys. Rev. D. {\bf 70}, 023006 (2004).
  
\bibitem{Collaboration:2009nf}
  The IceCube Collaboration,
  arXiv:0902.0675 [astro-ph.HE].

\bibitem{halzen}
  F.~Halzen,
  Eur.\ Phys.\ J.\  C {\bf 46}, 669 (2006)
  [arXiv:astro-ph/0602132].

\bibitem{gandhi}
  R.~Gandhi, C.~Quigg, M.~H.~Reno and I.~Sarcevic,
  Phys.\ Rev.\  D {\bf 58}, 093009 (1998)
  [arXiv:hep-ph/9807264].

\bibitem{nuArea}
  C.~Wiebusch and f.~t.~I.~Collaboration,
  arXiv:0907.2263 [astro-ph.IM].

\bibitem{bergstrom}
  L.~Bergstrom, J.~Edsjo and M.~Kamionkowski,
  Astropart.\ Phys.\  {\bf 7}, 147 (1997)
  [arXiv:astro-ph/9702037].

\bibitem{fgst}
See talks at TeV Particle Astrophysics 2009, July 13-19, 2009, http://www-conf.slac.stanford.edu/tevpa09/


\end{thebibliography}
\end{document}